\documentclass[11pt,english]{article}
\usepackage[T1]{fontenc}
\usepackage[latin9]{inputenc}
\usepackage{geometry}
\usepackage{color}
\usepackage{amsmath}
\usepackage{amssymb}
\usepackage{graphicx,url}
\usepackage{setspace}
\usepackage{times}
\usepackage[authoryear]{natbib}
\makeatletter

\@ifundefined{date}{}{\date{}}
\newtheorem{assumption}{Assumption}\newtheorem{theorem}{Theorem}\newtheorem{lemma}{Lemma}\newtheorem{remark}{Remark}

\newcommand*{\nindep}{%
	\mathbin{%                   % The final symbol is a binary math operator
		\mathpalette{\@indep}{\not}% \mathpalette helps for the adaptation
	}%
}
\newcommand*{\@indep}[2]{%
	\sbox0{$#1\perp\m@th$}%        box 0 contains \perp symbol
	\sbox2{$#1=$}%                 box 2 for the height of =
	\sbox4{$#1\vcenter{}$}%        box 4 for the height of the math axis
	\rlap{\copy0}%                 first \perp
	\dimen@=\dimexpr\ht2-\ht4-.2pt\relax
	\kern\dimen@
	{#2}%
	\kern\dimen@
	\copy0 %                       second \perp
} 

\newcommand{\pr}{\mathrm{pr}}

\newcommand{\T}{\mathrm{\scriptscriptstyle T}}

\newcommand{\bigCI}{\mathrel{\text{\scalebox{1.07}{$\perp\mkern-10mu\perp$}}}}
\newcommand{\ind}{\mathrel{\text{\scalebox{1.07}{$\perp\mkern-10mu\perp$}}}}

\usepackage{tikz}
\usetikzlibrary{arrows,shapes.arrows,shapes.geometric,
	shapes.multipart,backgrounds,decorations.pathmorphing,positioning,fit,automata}
\tikzset{
	>=stealth',
	true/.style={
		rectangle,
		draw=black, very thick,
		text width=6.5em,
		minimum height=2em,
		text centered,
		fill=gray, opacity = 0.5},
	punkt/.style={
		rectangle,
		rounded corners,
		draw=black, very thick,
		text width=6.5em,
		minimum height=2em,
		text centered},
	est/.style={
		circle,
		draw=black, very thick,
		minimum height=2.3em,
		text centered},
	estblue/.style={
		circle,
		draw=blue, very thick,
		text centered},
	estgreen/.style={
		circle,
		draw=mygreen, very thick,
		text centered},
	estmyblue/.style={
		circle,
		draw=myblue, very thick,
		text centered},
	estred/.style={
		circle,
		draw=red, very thick,
		text centered},
	estw/.style={
		circle,
		draw=white, very thick,
		text centered},
	mytext/.style={
		rectangle,
		draw=white, very thick,
		text width=6.5em,
		minimum height=2em,
		text centered},
	empty/.style={
		rectangle,
		draw=white, very thick,
		text width=10em,
		minimum height=2em,
		text centered},
	myempty/.style={
		circle,
		draw=white, very thick,
		text centered},
	shade/.style={
		circle,
		draw=black, very thick, fill=gray!80,
		minimum height=2.3em,
		text centered},
	missing/.style={
		circle,
		draw=black, very thick, fill=gray!20,
		text centered},
	weight/.style={
		circle,
		draw=black, very thick,
		text width=6.5em,
		minimum height=2em,
		text centered},
	pil/.style={
		->,
		thick,
		shorten <=2pt,
		shorten >=2pt,},
	piltx/.style={
		->,
		ultra	thick,
		shorten <=5pt,
		shorten >=5pt,},
	pilred/.style={
		->,
		thick,red,
		shorten <=2pt,
		shorten >=2pt,},
	pilblue/.style={
		->,
		thick,blue,
		shorten <=2pt,
		shorten >=2pt,},
	pilgreen/.style={
		->,
		thick,green,
		shorten <=2pt,
		shorten >=2pt,},
	pilgray/.style={
		->,
		thick,lightgray,
		shorten <=2pt,
		shorten >=2pt,},
	double/.style={
		<->,
		thick,
		shorten <=2pt,
		shorten >=2pt,},
	dash/.style={
		dashed,
		thick,
		shorten <=2pt,
		shorten >=2pt,},
	dashdouble/.style={
		<->,
		dashed,
		thick,
		shorten <=2pt,
		shorten >=2pt,}
}
\usepackage{babel}

\makeatother

\usepackage{subfigure}

\begin{document}
	
		%\markboth{D. Kong, S. Yang  and L. Wang}{Multi-cause causal inference}
	
	\title{\textbf{\huge{}Identifiability of causal effects with multiple causes and a binary outcome}}
	 \author{Dehan Kong\\
  Department of Statistical Sciences, University of Toronto\\[10pt]
  Shu Yang\\
  Department of Statistics, North Carolina State University\\[10pt]
  Linbo Wang \\
  Department of Statistical Sciences, University of Toronto \\[10pt]
    }
	\maketitle

%% The left and right page headers are defined here:
%\markboth{D. Kong, S. Yang and L. Wang}{Miscellanea}
%
%%% Here are the title, author names and addresses
%\title{Multi-cause causal inference with unmeasured confounding and binary outcome}
%
%\author{Dehan Kong}
%\affil{Department of Statistical Sciences, University of Toronto, Toronto, Ontario M5S 3G3, Canada  \email{kongdehan@utstat.toronto.edu}}
%
%\author{Shu Yang}
%\affil{Department of Statistics, North Carolina State University, Raleigh, North Carolina 27695, U.S.A.
%\email{syang24@ncsu.edu}}
%
%
%\author{\and Linbo Wang}
%\affil{Department of Statistical Sciences, University of Toronto, Toronto, Ontario M5S 3G3, Canada   \email{
%linbo.wang@utoronto.ca}}

%\author{for the Alzheimer's Disease
%Neuroimaging Initiative}

\maketitle

\begin{abstract}
Unobserved confounding presents a major threat to causal inference
from observational studies. Recently, several authors suggest that this
problem may be overcome in a shared confounding setting where multiple
treatments are independent given a common latent confounder. It has
been shown that under a linear Gaussian model for the treatments, the causal effect is not identifiable without parametric assumptions on the outcome model. In this paper,  we show that the causal effect is indeed identifiable if we  assume  a  general binary choice model for the outcome with a non-probit link. Our identification approach is based on  {the incongruence between Gaussianity of the treatments and latent confounder,  and
	non-Gaussianity} of a latent outcome variable.  We further develop
a two-step likelihood-based estimation procedure. 
% Our
% method is illustrated through simulations and a  real data application studying the
% causal relationship between the volumes of various brain regions and
% cognitive scores. 
\end{abstract}


\section{Introduction}

Unmeasured confounding presents a major challenge to causal inference
from observational studies. Without further assumptions, it is often
impossible to identify the causal effects of interest. Classical approaches
to mitigate bias due to unmeasured confounding include instrumental
variable methods \citep[e.g.,][]{angrist1996identification,hernan2006instruments,wang2018bounded},
causal structure learning \citep[e.g.,][]{drton2017structure}, invariance
prediction \citep[e.g.,][]{peters2016causal}, % regression continuity designs \citep[e.g.,][]{imbens2008regression}, 
negative controls \citep[e.g.,][]{kuroki2014measurement,miao2018identifying}
and sensitivity analysis \citep[e.g.,][]{cornfield1959smoking}. %parametric assumptions (cite LINGAMs and related works) 

In a recent stream of literature, several authors suggest an alternative
approach to this problem by assuming shared confounding between multiple
treatments and independence of treatments given the confounder \citep{wang2019blessings,tran2017implicit,ranganath2018multiple,wang2019multiple}.
These approaches leverage information in a potentially
high-dimensional treatment to aid causal identification. Such settings
are prevalent in many contemporary applications such as genetics,
recommendation systems and neuroimaging studies. Unfortunately, in
general the shared confounding structure is not sufficient for causal
identification. \citet[][Theorem 1]{d2019multi} show that %even 
under a linear Gaussian treatment model, except in trivial cases, the causal effects are not identifiable without parametric assumptions on the outcome model.   To address
this non-identifiability problem, \citet{d2019multi} and \cite{imai2019discussion} suggest collecting auxiliary variables such as negative controls or instrumental variables. Along this line, \citet{wang2019multiple}
show that the deconfounder algorithm of \citet{wang2019blessings}
is valid given a set of negative controls, and \citet{veitch2019using}
further find a negative control in network settings.

In this paper, we contribute to this discussion by 
establishing a new identifiability result of causal effects 
%providing a novel identification result 
assuming a general binary choice outcome model with a non-probit link, in addition to a linear Gaussian treatment model. Our result provides a counterpart to the non-identifiability result by  \citet[][Theorem 1]{d2019multi}.
We use parametric assumptions in place of auxiliary  data for causal identification. This is similar in spirit to  Heckman's selection model \citep{heckman1979sample}  for correcting bias from non-ignorable missing data. 
%To the best of our knowledge, this result is the first in the literature that requires no external data but still identifies the causal effect in this setting. 
In contrast to the case with normally-distributed treatments and outcome,
in general the observed data distribution may contain information
beyond the first two moments, thereby providing many more non-trivial
constraints for causal identification \citep{bentler1983simultaneous,bollen2014structural}.
% One of main advantages of the proposed method is that it requires
% no external data but still identifies the causal effect in this setting.
In particular, our approach leverages the incongruence between  Gaussianity of the treatments and latent confounder, and  non-Gaussianity of  a latent outcome variable to get causal identification.
A referee suggests that this is related to previous results by  \cite{peters2009detecting} and \citet[][\S 2.1]{imai2019discussion} in other contexts of causal inference.
%In particular, we exploit the symmetry and tail properties of the observed data distribution to identify the  causal effects. 
Our identification approach is accompanied by a simple likelihood-based
estimation procedure, and illustrations through synthetic and real
data analyses in the Supplementary Material.

%We first provide identifiability conditions for latent factor models
%for multiple continuous treatments and binary outcome without any
%observed confounders. Then we extend these conditions to the models
%with observed confounders. We show that incorporating additional confounders
%may aid in identifying model parameters. We then 
%propose a two-step likelihood-based estimation procedure for estimating the model parameters
%and causal effects, which is computationally efficient and theoretically 
% sound. Our proposed procedures are demonstrated via simulations and a real data application studying the causal relationship between
%the volume of different brain regions and the dementia behaviour. % This holds the potential to be the very first study of its kind investigating the causal relationship between neuroimaging data and human behaviour using observational data.

%\textcolor{red}{Is it possible to extend to general outcome models
%not restricted to logistic regression?}

\section{Framework}

Let $A=(A^{(1)},A^{(2)},\ldots,A^{(p)})^{\T}$ be a $p$-vector of continuous
treatments, $Y$ be an outcome and $X$ be a $q$-vector of observed
pre-treatment variables. The observed data $\{(X_{i},A_{i},Y_{i}):i=1,\ldots,n\}$
are independent samples from a super-population. Under the potential
 outcomes framework,
%  \citep{neyman1923applications,rubin1974estimating},
$Y(a)$ is the potential outcome if the patient had received treatment $a=(a^{(1)},\ldots,a^{(p)})^{\T}$.
We are interested in identifying and estimating the mean potential outcome $E\{Y(a)\}$. We make
the stable unit treatment value assumption under which $Y(a)$ is
well-defined and $Y=Y(a)$ if $A=a$.

\begin{figure}[!htbp]
	\centering
	\scalebox{0.8}{
	\begin{tikzpicture}[->,>=stealth',shorten >=1pt,auto,node distance=2.3cm,
	semithick, scale=0.50]
	pre/.style={-,>=stealth,semithick,blue,line width = 1pt}]
	\tikzstyle{every state}=[fill=none,draw=black,text=black]
	\node[shade] (U)                    {$U$};
	\node[est] (A1) [below left = 1.3cm and 2.3cm of U ] {$A^{(1)}$};
	\node[est] (A2) [below left = 1.3cm and 0.5cm of U ] {$A^{(2)}$};
	\node[myempty] (A3) [below right = 1.3cm and -0.6cm of U ] {$\ldots$};
	\node[est] (Ap) [below right = 1.3cm and 0.5cm of U ] {$A^{(p)}$};
	\node[est] (Y) [below right = 1.3cm and 2.3 cm of U ] {$Y(a)$};
	\node[est] (X) [below =  3cm of U] {$X$};
	\path   (U) edge node {} (A1)
	(U) edge node {} (A2)
	(U) edge node {} (Ap)
	(U) edge node {} (Y)
	(X) edge node {} (A1)
		(X) edge node {} (A2)
			(X) edge node {} (Ap)
				(X) edge node {} (Y);
	\end{tikzpicture}}
	\caption{A graphical illustration of the shared confounding setting. The latent ignorability assumption is encoded by the absence of arrows between $A^{(j)}$ and $Y(a)$ for $ j=1, \ldots, p $. The gray node denotes that $U$ is unobserved. }
	\label{fig:dag}
\end{figure}
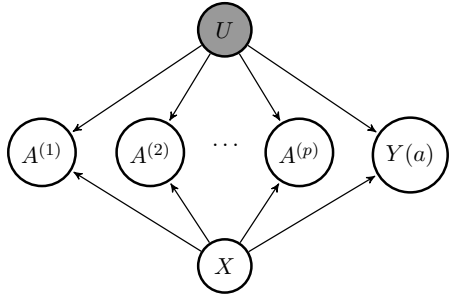

We assume the shared confounding structure under which the treatments are conditionally independent given the baseline covariates $X$ and  a scalar latent confounder $U$. Figure \ref{fig:dag} provides a graphical
illustration. \begin{assumption}[Latent ignorablity]\label{proposition:latent ign}For
	all $a$, $A\bigCI Y(a)\mid(X,U)$.
\end{assumption}
Under Assumption \ref{proposition:latent ign}, we have 
\begin{equation}
E\{Y(a)\}=E_{X,U}\{E(Y|A=a,X,U)\}.\label{eqn:gformula}
\end{equation}
We consider a latent factor model for the treatments:
\begin{equation}
U\sim\mathcal{N}(0,1),\qquad A=\theta U+\epsilon_{A},\label{model:1}
\end{equation}
where $\epsilon_{A}\sim\mathcal{N}\{0,\mathrm{diag}(\sigma_{A,1}^{2},\ldots,\sigma_{A,p}^{2})\}$
and $\epsilon_{A}\ind U$. \citet{wang2019blessings} suggest first
constructing an estimate of $U$, the so-called de-confounder, and
then use \eqref{eqn:gformula} to identify the mean potential outcomes
and causal contrasts. %\citet{wang2019blessings} was the first to establish Proposition
%\ref{proposition:latent ign}, which implies that the average potential
%outcome can be identified through that 
%for $a\in\mathcal{A}$; i.e., one can identify the causal effects
%even without observing the common confounders, and any latent variable
%$U$ that de-correlates the multiple treatments can be used as the
%replacement. The latent variable $U$ is coined as a deconfounder.
%To predict $U_{i}$ for each subject, \citet{wang2019blessings} assumed
%Under certain conditions, a consistent estimator $\widehat{U}_{i}$ of
%$U_{i}$ is used in combination with $X_{i}$ to adjust for confounding.
%However, the validity of their proposed de-confounding procedure is
%questioned due to lack of identification of model parameters in general.
%Without identifiability, the subsequent causal analysis is meaningless.
%We illustrate this issue using a simple counterexample.
However, as pointed out by \citet{d2019multi}, Assumption \ref{proposition:latent ign} and model \eqref{model:1} are not sufficient for identification of $E\{Y(a)\}$. See also Example S1 in the Supplementary
Material for a counterexample where $Y$ follows a Gaussian
structural equation model.

\section{Identification  with a binary outcome}

% \subsection{Identification  without observed confounders}

\label{sec:main}

We now study the identification  problem with a binary
$Y$, thereby operating under a different set of assumptions than
Example S1. %establish a formal framework
%for identification and estimation of causal effects under the latent
%factor models. 
To fix ideas, we first consider the case without measured covariates
$X$ and later extend these results to the case with $X$. We assume that treatments
$A$ follow the latent factor model \eqref{model:1}. %\textcolor{red}{although our result can be extended to factor models
%with multiple latent factors. \{I made up this, is this possible?\}}
%	If this assumption
%does not hold, one can simply let $\theta^{*}=-\theta$ and $U^{*}=-U$,
%and the positivity assumption on $\theta_{j}$ is satisfied for the
%model $A=\theta^{*}U^{*}+\epsilon_{A}$. For the binary outcome $Y(a)\in\{0,1\}$,
We also assume the following binary choice model: 
\begin{equation}
Y=1(T\leq\alpha+\beta^{\T}A+\gamma U),\label{model:binary:choice}
\end{equation}
where an auxiliary latent variable $T$, independent of $(A,U)$,
has a known cumulative distribution function $G$. Equivalently,
model \eqref{model:binary:choice} can be written as 
$
\pr(Y=1\mid A,U)=G(\alpha+\beta^{\T}A+\gamma U).
$
This class of models is general and includes common models for the
binary outcome. For example, when $T$ follows a logistic distribution
with mean zero and scale one, model \eqref{model:binary:choice} becomes
a logistic model; when $T$ follows a standard
normal distribution, model \eqref{model:binary:choice} is a probit
model; when $T$ follows a central-$t$ distribution, model \eqref{model:binary:choice}
is a robit  model \citep{liu2004robit, ding2014bayesian}.

Our main identification result  is summarized in Theorem \ref{Thm:1}. 

\begin{theorem}\label{Thm:1}Assume that Assumption \ref{proposition:latent ign},
	models \eqref{model:1}, \eqref{model:binary:choice} and the following
	conditions hold: 
	\begin{itemize}
		\item[(A1)] There exist at least three elements of $\theta=(\theta_{1},\ldots,\theta_{p})^{\T}$
		that are non-zero, and there exists at least one $j\in \{1,\ldots, p\}$ such that  $\gamma\theta_{j}  \neq 0$ and its sign is known a priori.
		\item[(A2)] $\pr(Y=1\mid A=a)$ is not a constant function of $a$. 
	\end{itemize}
	Then the parameters $\theta$, $\Sigma_{AA}$, $\alpha$, $\beta$,
	$\gamma$ and hence $E\{Y(a)\}$ are identifiable if and only if $T$
	is not deterministic or normally distributed. 
	
	%does not follow a normal distribution or  is not a constant almost surely.  %Therefore, for $a\in\mathcal{A}$,
	%$\mu(a)=E\{E(Y|A=a,U)\}=E\{g(\alpha+\beta^{\T}a+\gamma U)\}$ is identifiable.
\end{theorem}

Theorem \ref{Thm:1} entails that  identifiability of causal effects is guaranteed
as long as the outcome follows a non-trivial binary choice model with any link function other than the probit.
Condition (A1) is plausible when the latent confounder $U$ affects at
least three treatments, and for at least one of which, subject-specific
knowledge allows one to determine the signs of $\theta_{j}$ and  $\gamma$. Condition (A2) requires that the observed outcome means
differ across treatment levels, and can be checked from the
observed data.

We now present an outline of our identification strategy leading to Theorem \ref{Thm:1}. 
We first note that under model \eqref{model:1}, $(U,A^{\T})^{\T}$ follows
a joint multivariate normal distribution 
\begin{eqnarray}
\label{eqn:ua}
	\left(\begin{array}{c}
		U\\
		A
	\end{array}\right)\sim\mathcal{N}_{p+1}(0,\Sigma_{J}), &  & \Sigma_{J}=\left(\begin{array}{cc}
		1 & \theta^{\T}\\
		\theta & \Sigma_{AA}
	\end{array}\right),
\end{eqnarray}
where $\Sigma_{AA}=\theta\theta^{\T}+{\rm diag}(\sigma_{A,1}^{2},\ldots,\sigma_{A,p}^{2})$.
Therefore, $U|A$ follows a univariate normal distribution with
mean $\mu_{U|A}=\theta^{\T}\Sigma_{AA}^{-1}A$ and variance $\sigma_{U|A}^{2}=1-\theta^{\T}\Sigma_{AA}^{-1}\theta$. 

The starting point for our identification approach is the following orthogonalization of  $(U,A^{\T})^{\T}$.
Let $Z = (U-\mu_{U|A})/\sigma_{U|A}$ be the %normalized
standardized
latent confounder conditional on $A$. Then $Z \ind A$ and $Z$ follows a standard normal distribution. Model \eqref{model:binary:choice} then implies that
\begin{equation}
\label{eqn:7} Y{=}1(T\leq c_{1}+c_{2}^{\T}A+c_{3}Z),
\end{equation}
 where $c_{1}=\alpha$, $c_{2}=(c_{2}^{(1)},\ldots,c_{2}^{(p)})^{\T}=\beta+\gamma \theta^{\T} \Sigma_{AA}^{-1},$
$c_{3}=\gamma\sigma_{U|A}$ and $(A,T,Z)$ are  jointly independent.

The unknown parameters can then be identified in three steps. In the first step, we prove the identifiability of $\theta$ and $\Sigma_{AA}$ using standard results from factor analysis \citep{anderson1956statistical}. In the second step, we study  the binary choice model \eqref{eqn:7}, and show that both $c_2$ and the distribution of $T - c_1 - c_3 Z$ are identifiable up to a positive scale parameter. In the third step, we show that when the distribution of $T$ is non-deterministic and non-Gaussian, one can leverage the incongruence between Gaussianity of $Z$ and non-Gaussianity of $T$ to identify $c_1, c_3$ and  the scale parameter in the second step. The key to this step is the following Lemma \ref{keylemma}. Finally, we identify $\alpha, \beta, \gamma$ and hence $E\{Y(a)\}$ from $c_1, c_2, c_3, \theta, \Sigma_{AA}.$

\begin{lemma}\label{keylemma} Suppose $T_1=T-c_{1}-c_{3}Z$
	 and $T$ is independent of $Z$, where $Z$ follows a standard normal distribution, $c_{1}$ and $c_{3}$ are constants. The following statements are equivalent:
	\begin{itemize}
		\item[(I)]  There exist $(\widetilde{C},\widetilde{c_{1}},|\widetilde{c_{3}}|) \neq  (C,c_1,|c_3|)$, $\widetilde{T} \overset{\mathcal{D}}{=} T, \widetilde{Z} \overset{\mathcal{D}}{=} Z,$
		such that $C\widetilde{C}>0$, $\widetilde{T} \ind \widetilde{Z}$ and $CT_1 \overset{\mathcal{D}}{=} \widetilde{C}(\widetilde{T}-\widetilde{c_{1}}-\widetilde{c_{3}}\widetilde{Z}),$ where $E \overset{\mathcal{D}}{=} F$ denotes that random variables $E$ and $F$ have the same distribution;
		\item[(II)] $T$ is either deterministic or normally distributed.
	\end{itemize}
\end{lemma}

\begin{remark}
In our paper, we only allow $U$ to be a scalar. In this case, $\theta$ is identified up to sign flip from the factor model, and it may be possible to identify the sign of $ \theta$ from subject-matter knowledge. However, if $U$ is a multi-dimensional vector, then the factor model \eqref{model:1} becomes $ A=\Theta U+\epsilon_A $, where $\Theta$ is the loading matrix. In this case, $\Theta$ is only identifiable up to a rotation. Consequently, in general, there are infinitely many causal effect parameters that are compatible with the observed data distribution; see \cite{miao2020identifying} for related discussions.
\end{remark}

\begin{remark}
  Example S1 in the Supplementary Material shows that when the continuous outcome $Y$ follows a Gaussian structural model, $E\{Y(a)\}$ is not identifiable.  Intuitively, the binary
  outcome in a  probit regression can be obtained by  dichotomizing a continuous outcome following  a Gaussian distribution, and there is no reason to believe that dichotomization improves identifiability. So it should not be surprising that $E\{Y(a)\}$ is not identifiable in the probit case.

% It is worth noting that the non-identifiability under the Gaussian/probit case has to do with the fact that the Gaussian distribution is fully characterized by its first two moments, limiting the amount of information contained in the observed data distribution. Once we move away from the Gaussian/probit case, identifiability becomes easier. 
\end{remark}

In the presence of baseline covariates $X$, we
assume
\begin{eqnarray}
A & = & \theta U+BX+\epsilon_{A},\label{model:1:confounder}\\
\pr\{Y(a)=1|U,X\} & = & G(\alpha+\beta^{\T}a+\gamma U+\eta^{\T}X),\label{model:binary:choice:confounder}
\end{eqnarray}
where $X\ind(U,\epsilon_{A})$. 
% {\color{red} It looks like the condition can not be relaxed to uncorrelated in this case. Do we need to mention projection any more? } 
We also assume that% $ (U, A)\mid X $ follows a multivariate normal distribution conditional on $ X $:
\begin{equation}\label{conditionalnormality}
	\left.\left(\begin{array}{c}
		U \\
		A 
	\end{array}\right)\right| X
	\sim\mathcal{N}_{p+1}\left\{	
	\left(\begin{array}{c}
		0 \\
		BX 
	\end{array}\right)
	,\Sigma_{J}^{*}\right\},  
\Sigma_{J}^{*}=\left(\begin{array}{cc}
		1 &  \theta^{\T}\\
		\theta & \Sigma_{A\mid X} 
	\end{array}\right),
\end{equation}
where $ \Sigma_{A\mid X} = \Sigma_{AA}-B\Sigma_{XX}B^{\T}$ with $ \Sigma_{AA}$ and $ \Sigma_{XX}$ the covariances of $ A $ and $ X$, respectively. 
Then $U|X=x,A=a$ follows a univariate normal distribution with
mean $\mu_{U|x,a}=\theta^{\T} \Sigma_{A\mid X}^{-1} (a-Bx)$
and variance $\sigma_{U|x,a}^{2}=1-\theta^{\T}\Sigma_{A\mid X}^{-1}\theta$. 
% Similar to \eqref{Qtrepresentation}, we have 
% \begin{flalign*}
% &\pr(Y=1\mid A=a,X=x)  =\int G(\alpha+\beta^{\T}a+\gamma u+\eta^{\T}x)f_{U|X,A}(u|x,a)\de u\\
% & =\int G\{\alpha+(\beta^{\T}+\gamma\theta^{\T}\Sigma_{A\mid X}^{-1})a+(\eta^{\T}-\gamma\theta^{\T}\Sigma_{A\mid X}^{-1}B)x
% +\gamma\sigma_{U|x,a}v\}\phi(v)\de v. %\numberthis\label{Qxtrepresentation}
% \end{flalign*}
Identifiability of $E\{Y(a)\}$ can then be obtained similarly as in Theorem \ref{Thm:1}, except that now we replace condition (A2) with the
following weaker condition: 
	\begin{itemize}
		\item[(A2{*})] $\pr(Y=1\mid A=a,X=x)$ depends on $a$, or $x$, or both. Furthermore, if
		$\pr(Y=1\mid A=a,X=x)$ only depends on a subset of $x$, say $\{x_{j_{1}},x_{j_{2}},\ldots,x_{j_{k}},1\leq j_{1}<\ldots<j_{k}\leq q\}$, 
		then at least one of $\{X_{j_{1}},X_{j_{2}},\ldots,X_{j_{k}}\}$ has full support in $\mathbb{R}$.
	\end{itemize}

\noindent \begin{theorem}
\label{thm:2} Assume that Assumption \ref{proposition:latent ign},
	models \eqref{model:1:confounder}, \eqref{model:binary:choice:confounder}, and conditions \eqref{conditionalnormality},
	 (A1), (A2{*}) hold.
	Then the parameters $\theta$, $\Sigma_{AA}$, $\alpha$, $\beta$,
	$\gamma$, $\eta$ and hence $E\{Y(a)\}$ are identifiable if and
	only if $T$ is not deterministic or normally distributed. 
	
\end{theorem} 

The proof of Theorem \ref{thm:2} is similar to that  of Theorem \ref{Thm:1} and hence omitted.

\section{Discussion}

In this note, we consider the setting investigated by \cite{wang2019blessings} and others, including conditional independence among multiple treatments, and a linear Gaussian treatment model. We provide necessary and sufficient conditions for identifiability of causal effects under a binary choice model for the outcome. 

When the causal effects are identifiable, one can use the following likelihood-based
procedure to estimate the model parameters. Asymptotic normality and
resulting inference procedures follow directly from standard M-estimation
theory.

%\begin{description}
%	\item [{{Step$\ 1.$}}] Find the maximum likelihood estimator $(\widehat{\theta},\widehat{\Sigma}_{AA})^{\T}$
%	using off-the-shelf packages for factor analysis, such as the \texttt{factanal}
%	function in \texttt{R}. 
%	\item [{{Step$\ 2.$}}] Find the maximum likelihood estimator $(\widehat{\alpha},\widehat{\beta}^{\T},\widehat{\gamma})^{\T}$
%	by maximizing the conditional likelihood $\prod_{i=1}^{n}[r_{i}(\alpha,\beta,\gamma)^{Y_{i}}\{1-r_{i}(\alpha,\beta,\gamma)\}^{1-Y_{i}}]$,
%	where $r_{i}(\alpha,\beta,\gamma)=\pr(Y=1\mid A=A_{i};\alpha,\beta,\gamma,\widehat{\theta},\widehat{\Sigma}_{AA})$.
%	Monte Carlo method can be used to approximate the integral in \eqref{Qtrepresentation}. 
%\end{description}

% A two-step estimation procedure proceeds as follows.
\begin{description}
	\item [{{Step$\ 1.$}}] Let  $A^*$ be the residual of a linear regression of $A$ on $X$.
	Obtain the maximum likelihood estimators $\widehat{\theta}$
	and $\widehat{\Sigma}_{A\mid X}$
	based on a factor analysis on $A^{*}$, using off-the-shelf packages such as the \texttt{factanal}
	function in \texttt{R}. When there are no observed confounders $ X $, one can use $ A $ instead of $ A^* $ and perform the factor analysis. 
	\item [{{Step$\ 2.$}}] Estimate $(\alpha,\beta^{\T},\gamma,\eta)$ by
	maximizing the conditional likelihood $\prod_{i=1}^{n}[\widetilde{r}_{i}(\alpha,\beta,\gamma,\eta)^{Y_{i}}\{1-\widetilde{r}_{i}(\alpha,\beta,\gamma,\eta)\}^{1-Y_{i}}]$,
	where $\widetilde{r}_{i}(\alpha,\beta,\gamma,\eta)=\pr(Y=1\mid A=A_{i},X=X_{i};\alpha,\beta,\gamma,\eta,\widehat{\theta},\widehat{\Sigma}_{A\mid X})$.
	%$L(\eta_{OX}\mid\widehat{\eta}_{AX})$,
	%where $L(\eta_{OX}\mid\eta_{AX})=\prod_{i=1}^{n}[Q(X_{i},A_i)^{Y_{i}}\{1-Q(X_{i},A_i)\}^{1-Y_{i}}]$.
	%By representation \eqref{Qtrepresentation}, $Q(a)$ involves integral
	%over $u$, and we use Monte Carlo simulation to approximate the integral.
\end{description}
%Similarly as the maximum likelihood estimation in no observed confounder case, to make the solution unique, we add a constraint such that $\theta_{\min \{j:\theta_j \neq 0\}}>0$. The sign of $ \gamma $ is then determined by the sign of $ \theta $. 

In the Supplementary Material, we also report numerical results from  synthetic data analyses and illustrations on real data sets. In a recent note, \cite{grimmer2020ive} show that the deconfounder algorithm by \cite{wang2019blessings} may not consistently outperform naive regression ignoring the unmeasured confounder when outcome and treatments follow Gaussian models. In comparison, our numerical results suggest that under our identification conditions, the likelihood-based estimates outperform  naive regression estimates. Furthermore,  these estimates exhibit some robustness to violations of the binary choice model specification. Nevertheless, we end with a cautionary note that our results show that identification of causal effects in the multi-cause setting requires additional parametric structural assumptions, including the linear Gaussian treatment model and the binary choice outcome model.

% \section*{Acknowledgements}
% The authors thank the editor, the associate editor and  referees for their helpful comments and suggestions. The authors also thank Jiaying Gu, Stanislav Volgushev and Ying Zhou for insightful discussions that improve our main identification theorem.  

% \section*{Supplementary material}
% \label{SM}
% Supplementary material  includes examples, simulation results, and two data illustrations. 

\section*{Appendix}

\subsection*{Proof of Theorem \ref{Thm:1}}

We shall use the following notation. Let $A^{(-1)}=(A^{(k)}: k\neq 1)\in \mathbb{R}^{p-1}$, and define $a^{(-1)}\in \mathbb{R}^{p-1}$ and $c_2^{(-1)}\in \mathbb{R}^{p-1}$  analogously. Also denote $A^{(-1,-j)}=(A^{(k)}: k\notin\{ 1, j\})\in \mathbb{R}^{p-2}$.

We first establish the identifiability results for
$\theta$ and $\Sigma_{AA}$. When $p\geq3$, by condition (A1),  there exist at
least three non-zero elements of $\theta=(\theta_{1},\ldots,\theta_{p})^{\T}$. By \citet[][Theorem 5.5]{anderson1956statistical},
one can identify $\theta$ up to sign and uniquely identify $\sigma_{A}^{2}$. 
% By condition (A1), there exists at least one $\theta_{j}\neq0$, $1\leq j\leq p$ such that $ \gamma\theta_j\neq 0$ and the sign of $\gamma\theta_j$ is known.
As  $U$ is latent with a symmetric distribution around zero, without loss of generality, we %may
assume we know $ \gamma>0$ so that the sign of $\theta_j$ in condition (A1) is determined accordingly; if otherwise, we  %may 
redefine $U$ to its negative, and all the assumptions in Theorem \ref{Thm:1} hold if we also redefine $\theta_j$ and $\gamma$ to their negatives, respectively. It follows that both $\theta$ and $\Sigma_{AA}$ are identifiable.

We now study the binary choice model  \eqref{eqn:7}. This is a non-traditional binary-choice model as the right hand side of the inequality involves a latent variable $Z$. We hence let 
% \begin{equation}
    % \label{eqn:w}
    $T_1=T-c_{1}-c_{3}Z$
% \end{equation}
 so that $A\ind T_1$ and model \eqref{eqn:7} becomes
\begin{equation}
\label{eqn:8}
    Y{=}1(T_1\leq c_{2}^{\T}A).
\end{equation}
This is a binary choice model  first introduced in economics \citep[e.g.][]{cosslett1983distribution, gu2020nonparametric} and recently studied in statistics \citep[e.g.][]{tchetgen2018discrete}. Condition (A2) implies that there exists $j$ such that $c_{2}^{(j)}\neq0$. Without loss of generality, we assume $c_2^{(1)} \neq 0$. 
% The rest of the proof consists of two parts.
% \begin{enumerate}
%     \item[Claim 1] One can identify the sign of $c_2^{(1)}$,  the distribution of $T_1/c_2^{(1)}$  and  $c_2/ c_2^{(1)}$  from eqn. \eqref{eqn:8} and observed data on $A, Y$;
%     \item[Claim 2] One can identify $c_1, c_2^{(1)},  c_3$ from \eqref{eqn:w}, the distributions of $T_1/c_2^{(1)}, T, Z$  and the sign of $c_2^{(1)}.$
% \end{enumerate}

To identify the sign of $c_2^{(1)}$ and the distribution of $T_1/c_2^{(1)}$, note that \eqref{eqn:8} implies that 
\begin{equation}
    \label{eqn:mono}
    \pr (Y=1\mid A=a) = \pr (T_1 \leq c_2^{\T} A \mid A=a)  = \pr (T_1 \leq c_2^{\T} a),
\end{equation}
where the second equality holds since $A \ind T_1.$  Since $A$ follows a multivariate Gaussian distribution, \eqref{eqn:mono} holds for any $a\in \mathbb{R}^p.$
 Setting $a^{(-1)}= 0$ in \eqref{eqn:mono}, we can identify  $\pr(T_1 \leq c_2^{(1)} a^{(1)})$ for any $a^{(1)} \in \mathbb{R}$. Condition (A2) and \eqref{eqn:mono} guarantees that this is a monotone non-constant function of $a^{(1)}.$ It is easy to see that  $c_2^{(1)} > 0$ if and only if $\pr(T_1 \leq c_2^{(1)} a^{(1)})$ is an increasing function of $a^{(1)}$ so that the sign of $c_2^{(1)}$ is identifiable. Thus the distribution of $T_1/c_2^{(1)}$ is  identifiable.

% Fixing $a^{(-1)} \equiv \{a^{(k)}: k\neq 1\}$, $\pr (T_1 \leq c_2^{\T} a)$ is monotone increasing in and not a constant function of $c_2^{(1)} a^{(1)}$; hence $\pr (Y=1\mid A=a) $ is monotone increasing in $a^{(1)}$ if $c_2^{(1)}$ is positive, and monotone decreasing in $a^{(1)}$ if $c_2^{(1)}$ is negative. 

We now show that $c_2/ c_2^{(1)}$ is identifiable. Without loss of generality, we assume $c_2^{(1)}>0.$
If we let
$T_2=\left[T_1 - \left\{c_2^{(-1)}\right\}^{\T} A^{(-1)}\right]/{c_2^{(1)}},$ then 
 \eqref{eqn:mono} implies that for any $a^{(-1)} \in \mathbb{R}^{p-1},$
% \begin{equation}
%     \label{eqn:t2}
%     \pr(Y=1\mid A=a) = 
%     % \pr \left(  T_2 \leq a^{(1)} + \widetilde{c_2^{(-1)}} a^{(-1)}  \right) = 
%     \pr \left( \left. T_2 \leq a^{(1)} + \widetilde{c_2^{(-1)}} A^{(-1)} \right| A^{(-1)} = a^{(-1)} \right).
% \end{equation}
$$
    \pr(Y=1\mid A=a) = \pr \left( \left. T_2 \leq A^{(1)}  \right| A=a \right) = \pr \left( \left. T_2 \leq a^{(1)}  \right| A^{(-1)} = a^{(-1)} \right), \ \forall a^{(1)} \in \mathbb{R}.
$$
Consequently, the distribution, and hence the expectation, of $T_2\mid A^{(-1)} = a^{(-1)}$ is identifiable.  It follows that  for $j=2,\ldots, p$, we can also identify 
$$
   c_2^{(j)} / c_2^{(1)} = E[T_2\mid A^{(-1)} = 0] - E[T_2\mid A^{(-1,-j)} = 0, A^{(j)} = 1],
$$
where the equality holds since $A\ind T_1.$

We now turn to the third step in the proof.  Lemma \ref{keylemma} implies that $c_2^{(1)}, c_1, c_3^2$ are all identifiable if and only if $T$ is not deterministic or normally distributed. The sign of $c_3 = \gamma \sigma_{U\mid A}$ can then be determined from the sign of $\gamma,$ as $\sigma_{U\mid A} \geq 0.$ Thus, the parameters $\theta$, $\Sigma_{AA}$, $\alpha$, $\beta$, $\gamma$ and hence $E\{Y(a)\}$ are identifiable if and only if $T$
	is not deterministic or normally distributed, which finishes the proof.

\subsection*{Proof of Lemma \ref{keylemma}}

Without loss of generality, we assume $C = 1$. Let $\widetilde{T_1} = \widetilde{T}-\widetilde{c_1}-\widetilde{c_3}\widetilde{Z}$.

We first show that (II) implies (I). Suppose  $T\sim\mathcal{N}(\mu_{T},\sigma_{T}^{2})$, where $\sigma_T^2>0$ if $T$ is normally distributed, and $\sigma_T^2=0$ if $T$ is deterministic.
Then $T_1\sim\mathcal{N}(\mu_{T}-c_{1},\sigma_{T}^{2}+c_{3}^{2})$
and $\widetilde{C}\widetilde{T_1}\sim\mathcal{N}\{\widetilde{C}(\mu_{T}-\widetilde{c_1}),\widetilde{C}^2(\sigma_{T}^{2}+\widetilde{c_3}^2)\}$. It is easy to verify that if $\widetilde{C} = 2, \widetilde{c_1} = (\mu_T + c_1)/2, \widetilde{c_3}^2 = c_3^2/4 - 3\sigma_T^2/4,$ then $CT_1 \overset{\mathcal{D}}{=} \widetilde{C}\widetilde{T_1}.$

% Meanwhile, we have $T-\widetilde{c_1}-\widetilde{c_3}Z\sim\mathcal{N}\{\mu_{T}-\widetilde{c_1},\sigma_{T}^{2}+(c_{3}^{2})^{*}\}$.
% If $\widetilde{C}T_1\overset{\mathcal{D}}{=}T-\widetilde{c_1}-\widetilde{c_3}Z$,
% one can obtain $\widetilde{c_1}=\widetilde{C}c_{1}-(\widetilde{C}-1)\mu_{T}$ and $(\widetilde{c_3})^{2}=(\widetilde{C}^{2}-1)\sigma_{T}^{2}+\widetilde{C}^{2}c_{3}^{2}$
% for any constant $\widetilde{C}>0$, $\widetilde{C}\neq1$ such that $(\widetilde{C}^{2}-1)\sigma_{T}^{2}+\widetilde{C}^{2}c_{3}^{2}\geq0$.
% Therefore, there exist non-unique sets of $(\widetilde{C},\widetilde{c_1},\widetilde{c_3})$
% satisfying $\widetilde{C}>0$ and (i) $\widetilde{C}>0$ and $\widetilde{C}\neq1$, or (ii) $\widetilde{c_1}\neq c_{1}$, or (iii) $(\widetilde{c_3})^{2}\neq c_{3}^{2}$, 
% such that $\widetilde{C}T_1\overset{\mathcal{D}}{=}$$T-\widetilde{c_1}-\widetilde{c_3}Z$.

% Suppose $T$ is deterministic so that $T$ is a constant
% almost surely. In this case, we can view $T$ as a degenerate normal
% with mean $\mu_{T}$ and variance $\sigma_{T}^{2}=0$. Similar as
% the previous argument, one can obtain $\widetilde{c_1}=\widetilde{C}c_{1}-(\widetilde{C}-1)\mu_{T}$
% and $(\widetilde{c_3})^{2}=\widetilde{C}^{2}c_{3}^{2}$ for any constant $\widetilde{C}>0$, $\widetilde{C}\neq1$. 

 We then show that (I) implies (II). 
% Assume there exists a different set of constants $(\widetilde{C},\widetilde{c_1},\widetilde{c_3})$ satisfying (i) $\widetilde{C}>0$ and $\widetilde{C}\neq1$, or (ii) $\widetilde{c_1}\neq c_{1}$, or (iii) $(\widetilde{c_3})^{2}\neq c_{3}^{2}$, 
% such that $\widetilde{C}T_1\overset{\mathcal{D}}{=}T-\widetilde{c_1}-\widetilde{c_3}Z$
% and $T_1\overset{\mathcal{D}}{=}T-c_{1}-c_{3}Z$. 
We start by showing that $\widetilde{C}\neq 1$. If otherwise,  $T-c_1-c_3Z\overset{\mathcal{D}}{=}\widetilde{T}-\widetilde{c_1}-\widetilde{c_3}\widetilde{Z}$. We then have for all $t\in \mathbb{R},$ $\phi_{T}(t)\phi_{c_1+c_{3}Z}(t)=\phi_{T}(t)\phi_{\widetilde{c_1}+\widetilde{c_3}Z}(t)$ and hence $\phi_{c_1+c_{3}Z}(t)=\phi_{\widetilde{c_1}+\widetilde{c_3}Z}(t)$, where $\phi_{T}(t)$ is the characteristic function of $T$. As a result, $c_1+c_{3}Z\overset{\mathcal{D}}{=}\widetilde{c_1}+\widetilde{c_3}Z$, which implies that $(c_1,|c_3|)=(\widetilde{c_1},|\widetilde{c_3}|)$. Contradiction! 

We now let  $c_1^*=\widetilde{C}\widetilde{c_1}$ and $c_3^*=\widetilde{C}\widetilde{c_3}$ so that
 $\widetilde{C}\widetilde{T}-c_1^*-c_3^*\widetilde{Z}\overset{\mathcal{D}}{=}T-c_{1}-c_{3}Z$.
We first consider the case where $|c_3^*|=|c_{3}|$. By a similar characteristic function argument as above,  $\widetilde{C}T-c_1^*\overset{\mathcal{D}}{=}T-c_{1}$, so $T$ is a constant almost surely. We next consider
 the case where $|c_3^*|\neq |c_{3}|$. Without
loss of generality, we assume $|c_3^*|>|c_{3}|$. By a similar characteristic function argument as above, we have that 
% $c_3^*\widetilde{Z}\overset{\mathcal{D}}{=}\mathcal{N}\{0,(c_3^*)^{2}\}\overset{\mathcal{D}}{=}\widetilde{Z}+c_{3}Z$,
% where $\widetilde{Z}\ind Z$, and $\widetilde{Z}\sim\mathcal{N}\{0,(c_3^*)^{2}-c_{3}^{2}\}$.
% Thus, we have $Ts-c_1^*-\widetilde{Z}-c_{3}Z\overset{\mathcal{D}}{=}T-c_{1}-c_{3}Z$.
% Let $A=Ts-c_1^*-\widetilde{Z}$ and $B=T-c_{1}$. As $T\ind Z$
% and $\widetilde{Z}\ind Z$, we have $\phi_{A}(t)\phi_{c_{3}Z}(t)=\phi_{B}(t)\phi_{c_{3}Z}(t)$
% for any $t\in\mathbb{R}$. As $\phi_{c_{3}Z}(t)\neq0$ for any $t\in\mathbb{R}$,
% we have $\phi_{A}(t)=\phi_{B}(t)$ for any $t\in\mathbb{R}$, i.e.,
% $A\overset{\mathcal{D}}{=}B$. Therefore, we have $T\overset{\mathcal{D}}{=}Ts+c_{1}-c_1^*-\widetilde{Z}$.
% By writing $V=c_{1}-c_1^*-\widetilde{Z}$, we have 
\begin{equation}
\label{eqn:t}
    T\overset{\mathcal{D}}{=}\widetilde{C}T+V,
\end{equation}
where $V\ind T$ and $V\sim\mathcal{N}(\mu_{V},\sigma_{V}^{2})$ with
$\mu_{V}=c_{1}-c_1^*$ and $\sigma_{V}^{2}=(c_3^*)^{2}-c_{3}^{2}$. 
Eqn. \eqref{eqn:t} implies that 
\begin{equation}
\phi_{T}(t)=\phi_{T}(\widetilde{C}t)\phi_{V}(t)=\phi_{T}(\widetilde{C}^2t)\phi_{V}(\widetilde{C}t)\phi_{V}(t)=\cdots=\phi_{T}(\widetilde{C}^{K}t)\mathop\prod\limits_{k=1}^{K}\phi_{V}(\widetilde{C}^{k-1}t) = \cdots.\label{chariteration}
\end{equation}
Consequently, 
\begin{equation}
T\overset{\mathcal{D}}{=}\widetilde{C}T+V_{1}\overset{\mathcal{D}}{=}\widetilde{C}(\widetilde{C}T+V_{2})+V_{1}\overset{\mathcal{D}}{=}\cdots\overset{\mathcal{D}}{=}\widetilde{C}^{K}T+\sum_{k=1}^{K}\widetilde{C}^{k-1}V_{k}\overset{\mathcal{D}}{=}\cdots,\label{iteration}
\end{equation}
where $V_{k},k=1,\ldots,K,\ldots$ are identically and independently distributed,
and are independent of $T$.
 We will now show $\widetilde{C}<1.$ If otherwise, we have $\widetilde{C}>1$. Let $\|\cdot\|$ denote the modulus of a complex number. For any $t>0$, by \eqref{chariteration} and the property of a normal distribution, $\|\phi_{T}(t)\|\leq \|\phi_{V}(\widetilde{C}^{K-1}t)\|\rightarrow0$
as $K\rightarrow\infty$. This is a contradiction as   by the continuity of  the characteristic function,  $\lim\limits_{t\rightarrow 0} \phi_T(t) = 1.$

% $\phi_{T}(0)=0$, which contradicts with the
% definition of the characteristic function as $\phi_{T}(0)$ is always
% $1$. Therefore, we have $s<1$. 

We can now see that in \eqref{iteration}, as $K\rightarrow\infty$,
 $\widetilde{C}^{K}T\rightarrow0$ in probability, and  $\sum_{k=1}^{K}\widetilde{C}^{k-1}V_{k}\rightarrow\mathcal{N}\{(1-\widetilde{C})^{-1}\mu_{V},(1-\widetilde{C}^{2})^{-1}\sigma_{V}^{2}\}$
in distribution. Therefore,  $T\sim\mathcal{N}\{(1-\widetilde{C})^{-1}\mu_{V},(1-\widetilde{C}^{2})^{-1}\sigma_{V}^{2}\}$. We have hence finished the proof.

\bibliographystyle{dcu}
\bibliography{multiple,causal}

\end{document}